\documentclass[twocolumn,twoside]{revtex4}
\usepackage{graphicx}
\usepackage{fancyhdr}
\def\beq{\begin{equation}}
\def\eeq{\end{equation}}
\def\bea{\begin{eqnarray}}
\def\eea{\end{eqnarray}}
\def\nn{\nonumber}
\def\roughly#1{\mathrel{\raise.3ex\hbox
{$#1$\kern-.75em\lower1ex\hbox{$\sim$}}}}

\def\sla#1{\raise.15ex\hbox{$/$}\kern-.57em #1}% Feynman slash

\def\BDstartaunu{{\bar B}^0 \to D^{*+} \tau^{-} {\bar\nu}_\tau}
\def\BDstarmunu{{\bar B}^0 \to D^{*+} \mu^{-} {\bar\nu}_\mu}
\def\bctaunu{b \to c \tau^- {\bar\nu}}
\def\bcmunu{b \to c \mu^- {\bar\nu}}

\def \cB{{\cal B}}
\def \cM {{\cal M}}
\def \cL{{\cal L}}
\def \cH{{\cal H}}
\def \cA {{\cal A}}
\def \SM{{\rm SM}}

\def \mAp{\mathcal{A}_+}
\def \mAm{\mathcal{A}_-}
\def \mAn{\mathcal{A}_{0}}
\def \mAt{\mathcal{A}_{t}}
\def \mAperp{\mathcal{A}_\perp}
\def \mApar{\mathcal{A}_\parallel}
\def \mAperpT{\mathcal{A}_{\perp, T}}

\def \mA0T{\mathcal{A}_{0, T}}

\begin{document}

\title{\boldmath CP Violation in ${\bar B}^0 \to D^{*+} \ell^-
  {\bar\nu}_\ell$}


\author{David London}
\affiliation{Physique des Particules, Universit\'e de Montr\'eal, \\
C.P. 6128, succ.\ centre-ville, Montr\'eal, QC, Canada H3C 3J7}

\begin{abstract}
At present, there are discrepancies with the predictions of the
standard model in ${\bar B}^0 \to D^{*+} \ell^- {\bar\nu}_\ell$
decays, hinting at the presence of new physics (NP) in $b \to c \tau^-
{\bar\nu}$. Various NP models have been proposed to explain the
data. In this talk, I discuss how the measurement of CP-violating
observables in ${\bar B}^0 \to D^{*+} \ell^- {\bar\nu}_\ell$ can be
used to differentiate the NP scenarios.
\end{abstract}

\maketitle

\thispagestyle{fancy}


\noindent
{\it This talk is based on work done in collaboration with
  B. Bhattacharya, A. Datta and S. Kamali
  \cite{Bhattacharya:2019olg}.}

\vskip1truemm

At the present time, there are discrepancies with the predictions of
the standard model (SM) in the measurements of $R_{D^{(*)}} \equiv
\cB(\bar{B} \to D^{(*)} \tau^{-} {\bar\nu}_\tau)/\cB(\bar{B} \to
D^{(*)} \ell^{-} {\bar\nu}_\ell)$ ($\ell = e,\mu$) and $R_{J/\psi}
\equiv \cB(B_c^+ \to J/\psi\tau^+\nu_\tau) / \cB(B_c^+ \to
J/\psi\mu^+\nu_\mu)$. The experimental results from before Moriond,
2019 are shown in Table I.  The deviation from the SM in $R_D$ and
$R_{D^*}$ (combined) was $\sim 3.8\sigma$ \cite{Abdesselam:2017kjf,
  Bernlochner:2017jka, Bigi:2017jbd,Jaiswal:2017rve}, while in
$R_{J/\psi}$ it is 1.7$\sigma$ \cite{Watanabe:2017mip}.

%%%%%%%%%%%%%%% [Table 1] %%%%%%%%%%%%%%%
\begin{table}[h]
\label{tab:obs_meas}
\begin{center}
\caption{Measured values of $\bctaunu$ observables from before Moriond, 2019.}
\begin{tabular}{|c|c|}
 \hline
Observable & Measurement/Constraint \\
\hline
$R_{D^*}^{\tau/\ell}/(R_{D^*}^{\tau/\ell})_\SM$ & $1.18 \pm 0.06$ \cite{RD_BaBar, RD_Belle, RD_LHCb,Abdesselam:2016xqt} \\
$R_{D}^{\tau/\ell}/(R_{D}^{\tau/\ell})_\SM$ & $1.36 \pm 0.15$ \cite{RD_BaBar, RD_Belle, RD_LHCb,Abdesselam:2016xqt} \\
$R_{D^*}^{\mu/e}/(R_{D^*}^{\mu/e})_\SM$ & $1.00 \pm 0.05$ \cite{Abdesselam:2017kjf} \\
$R_{J/\psi}^{\tau/\mu}/(R_{J/\psi}^{\tau/\mu})_\SM$ & $2.51 \pm 0.97$ \cite{Aaij:2017tyk} \\
 \hline
\end{tabular}
\end{center}
\end{table}
%%%%%%%%%%%%%%% [Table 1] %%%%%%%%%%%%%%%

\noindent
At Moriond, 2019, Belle announced new results \cite{Abdesselam:2019dgh}:
\bea
R_{D^*}^{\tau/\ell}/(R_{D^*}^{\tau/\ell})_\SM &=& 1.10 \pm 0.09 ~, \nn\\
R_{D}^{\tau/\ell}/(R_{D}^{\tau/\ell})_\SM &=& 1.03 \pm 0.13 ~.
\eea
These are in better agreement with the SM, so that the deviation from
the SM in $R_D$ and $R_{D^*}$ (combined) has been reduced from $\sim
3.8\sigma$ to $3.1\sigma$.

Even so, taken together, these measurements still hint at the presence
of new physics (NP) in $\bctaunu$ decays.

$\bctaunu$ is a charged-current process. The NP explanations that have
been examined include a $W^{\prime \pm}$, an $H^\pm$, or several
different types of leptoquarks (LQs). It was shown in
Ref.~\cite{Alonso:2016oyd} that considerations of the rate for $B_c^-
\to \tau^- {\bar\nu}$ disfavour NP models involving an $H^\pm$. Still,
this leaves a variety of different NP explanations. Assuming that NP
is indeed present in $\bctaunu$, how can we distinguish among these
possibilities? One idea is to use measurements of CP violation (CPV)
in $\BDstartaunu$ \cite{Bhattacharya:2019olg}.

The best-known CPV signal is direct CPV, in which the direct CP
asymmetry $A_{dir}$ is proportional to $\Gamma({\bar B}^0 \to D^{*+}
\tau^- {\bar\nu}_\tau) - \Gamma(B^0 \to D^{*-} \tau^+ \nu_\tau)$. Now,
CPV can only arise due to the interference of (at least) two
amplitudes with a relative weak (CP-odd) phase. But $A_{dir} \ne 0$
only if the interfering amplitudes also have different strong
(CP-even) phases. In $\BDstartaunu$, the only hadronic transition is
${\bar B} \to D^*$. This means that the strong phases of all
amplitudes, both SM and NP, are approximately equal, which then
implies that, even if $A_{dir}$ is nonzero, it is expected to be
small.

Instead, as we will see, the main CPV effects are CPV asymmetries in
the angular distribution of ${\bar B}^0 \to D^{*+} (\to D^0 \pi^+)
\tau^- {\bar\nu}_\tau$. Such asymmetries are a generalization of
triple-product asymmetries \cite{TPs, Gronau:2011cf,
  Duraisamy:2013kcw}, and are kinematical effects. That is, they can
be nonzero only if the interfering amplitudes have different Lorentz
structures. This allows us to distinguish different NP explanations.

Unfortunately, there is a practical problem. The $\BDstartaunu$
angular distribution requires the knowledge of the three-momentum
${\vec p}_\tau$. However, this cannot be measured, due to the missing
final $\nu_\tau$ in the decay of the $\tau^-$. A full analysis will
need to include information from the decay products of the $\tau$. My
collaborators and I are looking at this (it is work in progress), but
as a first step we examined the NP contributions to CPV angular
asymmetries in $\BDstarmunu$ \cite{Bhattacharya:2019olg}. Since ${\vec
  p}_\mu$ is measurable, this angular distribution can be
reconstructed. There are two reasons for starting with this
process. First, LHCb has announced \cite{Marangotto:2018pbs} that it
will perform a detailed angular analysis of this decay, with the aim
of extracting the coefficients of the CPV angular asymmetries. It is
therefore important to show exactly what the implications of these
measurements are for NP. Second, NP that contributes to $\bctaunu$ may
well also contribute to $\bcmunu$, leading to deviations from the SM
in $\BDstarmunu$.
 
Below, I sketch out the derivation of the angular distribution. For
all the details, the reader should consult
Ref.~\cite{Bhattacharya:2019olg}.

We begin by examining $\BDstarmunu$ within the SM.  The decay is
interpreted as ${\bar B}^0 \to D^{*+} (\to D^0 \pi^+) W^{*-} (\to
\mu^- {\bar\nu}_\mu)$, and the amplitude is written as
\beq
\cM_{(m;n)}(B\to D^* W^*) = \epsilon^{*\mu}_{D^*}(m) \, M_{\mu\nu} \, \epsilon^{*\nu}_{W^*}(n) ~. 
\eeq
Here, the (real) $D^{*+}$ has 3 polarizations, $m = +, -, 0$, while
the (virtual) $W^{*-}$ has 4 polarizations, $n = +, -, 0, t$ ($t =$
timelike).

Of the twelve $D^{*+}$-$W^{*-}$ polarization combinations, only four
are allowed by conservation of angular momentum: $++$, $--$, $00$,
$0t$. This implies that the decay is governed by four helicity
amplitudes: $\mAp$, $\mAm$, $\mAn$, $\mAt$. The decay amplitude can
then be written as
\bea
&& \cM(B\to D^*(\to D\pi)W^*(\to\mu^-{\bar\nu}_\mu)) \nn\\
&& ~~~~~ \propto
\sum\limits_{m=t,\pm,0} g_{mm} \, \cH_{D^*}(m) \, \cA_m \, \cL_{W^*}(m) ~,
\label{SMamp}
\eea
where $\cH_{D^*}$ and $\cL_{W^*}$ are, respectively, the hadronic and
leptonic matrix elements.  

We now add NP. There are two effects. First, we take $W^* \to N^*$,
where $N = S-P$ $(\equiv SP)$, $V-A$ $(\equiv VA)$, and $T$ represent
new interactions involving the left-handed neutrino. (Note that $VA$
includes the SM.) In the presence of these new interactions, there are
now more helicities. Previously, we had only $VA$, leading to the
helicity amplitudes $\mAp$, $\mAm$, $\mAn$, and $\mAt$. Now, there are
four more helicity amplitudes. The $SP$ interaction leads to
$\cA_{SP}$, while the $T$ interaction generates $\cA_{+,T}$,
$\cA_{0,T}$, and $\cA_{-,T}$.

Second, there are new contributions to the hadronic current:
\bea
{\cal H}_{eff} &=& \frac{G_F V_{cb}}{\sqrt{2}} \Bigl\{
\left[ g_S \, {\bar c} b + g_P \, {\bar c} \gamma_5 b \right] {\bar \ell} (1 - \gamma_5) \nu_\ell
\nn\\
&& \hskip-1truecm
+~\left[ (1 + g_L) \, {\bar c} \gamma_\mu (1 - \gamma_5) b + g_R \, {\bar c} \gamma_\mu (1 + \gamma_5) b \right] \nn\\
&& \hskip1truecm 
\times ~{\bar \ell} \gamma^\mu (1 - \gamma_5) \nu_\ell \nn\\
&& \hskip-1truecm
+~g_T \, {\bar c} \sigma^{\mu\nu} (1 - \gamma_5) b
{\bar \ell} \sigma_{\mu\nu} (1 - \gamma_5) \nu_\ell + h.c. \Bigr\} ~.
\eea

Including both SM and NP contributions, we now write
\bea
&& \cM(B\to D^*(\to D\pi)N^*(\to\mu^-{\bar\nu}_\mu)) \nn\\
&& \hskip 1.5truecm
= \cM^{SP} + \cM^{VA} + \cM^{T} ~,
\eea
where each term includes a sum over the relevant $D^*$ and $N^*$
helicities. The point is that, in the presence of NP, the amplitude now
contains a variety of Lorentz structures. (In the SM, we had only
$\cM^{VA}$ [Eq.~(\ref{SMamp})].)

We now compute $|\cM|^2$. This generates two types of terms: (i)
$|\cA_i|^2 f_i({\rm momenta})$ and (ii) the interference terms ${\rm
  Re}[\cA_i \cA_j^* f_{ij}({\rm momenta})]$. The momenta are defined
in Fig.~1. The computation of the quantities $f_i$ and $f_{ij}$ yields
the angular distribution.

\begin{figure}[h]
\centering
\includegraphics[width=80mm]{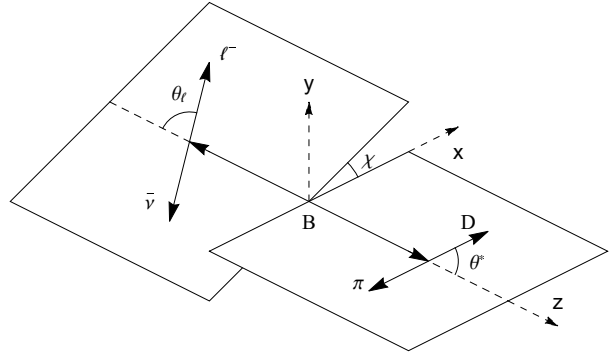}
\caption{Definition of the angles in the $\bar{B} \to D^* (\to D \pi) \ell^- \bar{\nu}_{\ell}$ distribution.} 
\label{fig:angdis}
\end{figure}

Here is the key point: in the interference terms, sometimes there is
an additional factor of $i$ in $f_{ij}({\rm momenta})$ (e.g., from
${\rm Tr}[\gamma_\mu \gamma_\nu \gamma_\rho \gamma_\sigma \gamma_5] =
4 i \epsilon_{\mu\nu\rho\sigma}$). In this case, the coefficient is
${\rm Im}[\cA_i \cA_j^*]$, which is sensitive to phase differences.

Furthermore, in this decay, the SM and NP strong phases are all
approximately equal. This implies that ${\rm Im}[\cA_i \cA_j^*]$
involves only the weak-phase difference. Such terms are therefore, by
themselves, signals of CP violation!

The complete angular distribution contains many CPV observables: some
are suppressed by $m_\mu^2/q^2$ or $m_\mu/\sqrt{q^2}$, and some are
unsuppressed. $q^2$ is typically of $O(m_b^2)$, so that the
suppression is significant. (On the other hand, if measurements can be
made in that region of phase space where $q^2 = O(m_\mu^2)$, here the
suppression is removed.) The unsuppressed observables are given in
Table II.

%%%%%%%%%%%%%%% [Table 2] %%%%%%%%%%%%%%%
\begin{table}[h]
\begin{center}
\caption{Unsuppressed CPV observables.}
\begin{tabular}{|l|l|}
\hline
Coefficient                                                                                        & Angular Function \\ \hline
${\rm Im}(\mAperp \mAn^*)$                        &      $- \sqrt{2} \sin 2\theta_\ell \sin 2\theta^* \sin\chi$               \\ \hline
$ {\rm Im}(\mApar \mAperp^*)$                     &     $ 2 \sin^2 \theta_\ell \sin^2 \theta^* \sin 2\chi$                \\ \hline
$ {\rm Im}(\mAn \mApar^*)$                        &      $- 2\sqrt{2} \sin\theta_\ell \sin 2\theta^* \sin\chi$               \\ \hline
$ {\rm Im}(\cA_{SP} \mAperpT^*)$             &       $- 8\sqrt{2} \sin\theta_\ell \sin 2\theta^*\sin\chi$              \\ \hline
\end{tabular}
\end{center}
\end{table}
%%%%%%%%%%%%%%% [Table 2] %%%%%%%%%%%%%%%

Which NP couplings are involved in these observables? ${\rm
  Im}(\mAperp \mAn^*)$, ${\rm Im}(\mApar \mAperp^*)$ and $ {\rm
  Im}(\mAn \mApar^*)$ are all generated by ${\rm
  Im}[(1+g_L+g_R)(1+g_L-g_R)^*]$, while $ {\rm Im}(\cA_{SP}
\mAperpT^*)$ is related to ${\rm Im}(g_Pg_T^*)$.

If the $\BDstarmunu$ angular distribution is measured, here is the NP
information that it yields:
\begin{itemize}

\item Most proposed NP models contribute only to $g_L$ (like the
  SM). If any CPV observables are found to be nonzero, these simple
  models are ruled out.

\item Suppose that the angular distribution contains, for example, a
  CPV $\sin 2\theta_\ell \sin 2\theta^* \sin\chi$ term. This implies
  that $g_R \ne 0$. In this case, one also expects to see nonzero
  $\sin^2 \theta_\ell \sin^2 \theta^* \sin 2\chi$ and $\sqrt{2}
  \sin\theta_\ell \sin 2\theta^* \sin\chi$ terms.

\item On the other hand, if the $\sin 2\theta_\ell \sin 2\theta^*
  \sin\chi$ term were found to vanish, this would imply that $g_R = 0$
  (or that its phase is the same as that of $(1 + g_L)$). In this
  case, the measurement of a nonzero $\sqrt{2} \sin\theta_\ell \sin
  2\theta^* \sin\chi$ term would imply that ${\rm Im}(g_Pg_T^*) \ne
  0$.

\item In all cases, additional information comes from the measurement
  of the CP-conserving pieces of the angular distribution. For
  example, both $ |\cA_{SP}|^2$ and $ |\mAperpT|^2$ can be determined
  from the angular distribution, so in principle we will know if they
  are nonzero (though we will have no information about their phases).

\item If measurements can be made in that region of phase space where
  $q^2 = O(m_\mu^2)$, removing the suppression factors $m_\mu^2/q^2$
  or $m_\mu/\sqrt{q^2}$ from some CPV observables, additional
  information can be obtained.

\end{itemize}

Another question is: what NP models can generate the new hadronic
couplings $g_R$, $g_P$, $g_T$?
\begin{enumerate}

\item The $R_2$ and $S_1$ LQ models generate $g_T$, while the $U_1$,
  $R_2$, $S_1$ and $V_2$ LQ models generate $g_P$. Thus, if ${\rm
  Im}(g_Pg_T^*) \ne 0$ is found, this points to a model containing two
  (different) LQs.

\item LQ models do not produce $g_R$. This coupling can arise, for
  example, in a model that includes both a $W'_L$ and a $W'_R$ that
  mix.

\end{enumerate}

Finally, I report on some work in progress. Earlier, it was argued
that a full analysis of the angular distribution of ${\bar B}^0 \to
D^{*+} (\to D^0 \pi^+) \tau^- {\bar\nu}_\tau$ must include information
from the decay products of the $\tau^-$. My collaborators and I have
looked at this, focusing on the decays $\tau^- \to \pi^- \nu_\tau$ and
$\tau^- \to \rho^- \nu_\tau$, with $\rho^- \to \pi^- \pi^0$ and $\pi^-
\pi^+ \pi^-$. When one takes into account the momenta of the decay
products of the $\tau^-$, there are now new angular observables, so
that we expect that the angular distributions using these $\tau^-$
decays will furnish complementary information to that obtained from
$\BDstarmunu$.  Our preliminary results confirm this. For example, in
$\BDstarmunu$, CPV terms proportional to ${\rm Im}(g_P (1 + g_L)^*)$
are suppressed by $m_\mu/\sqrt{q^2}$.  But in ${\bar B}^0 \to D^{*+}
\tau^- (\to \pi^- \nu_\tau) {\bar\nu}_\tau$, they are unsuppressed.

To summarize, the anomalies in $R_{D^{(*)}}$ and $R_{J/\psi}$ hint at
NP in $\bctaunu$. A variety of NP models have been proposed to explain
the data.  It has been suggested that these models can be
distinguished through the measurement of CP violation in
$\BDstartaunu$ \cite{Bhattacharya:2019olg}. In this talk, I have
described the first step, namely looking at the NP contributions to
CPV angular asymmetries in $\BDstarmunu$, which will be measured by
LHCb.

Our results can be summarized as follows:
\begin{enumerate}

\item Model-independent analysis: We allow for NP with new Lorentz
  structures. The interference of two contributions with different
  Lorentz structures leads to CP-violating angular asymmetries. We
  identify the CPV asymmetries in $\BDstarmunu$, and show how they
  depend on the NP parameters.

\item Model-dependent analysis: There are two classes of models that
  have been proposed to explain the data, involving a $W'$ or a LQ. In
  the simplest (most popular) models, the NP couples only to LH
  particles. If CPV is observed, these models are ruled out. We show
  how the other models can be distinguished, depending on which CPV
  asymmetries are found to be nonzero.

\end{enumerate}

\begin{acknowledgments}
I thank the FPCP2019 organizers for a wonderful event, my first
``live'' conference in over 15 years! This work was financially
supported in part by NSERC of Canada.
\end{acknowledgments}

\bigskip % extra skip inserted

\end{document}